# A little tour across the wonderful realm of meteor radiometry

**Jean-Louis Rault[1,2], François Colas[1]**

[1] IMCCE, Observatoire de Paris, France

[2] IMO Radio Commission, Hove, Belgium

Jean-Louis.Rault@obspm.fr

This paper describes the path strewn with pitfalls encountered during the development of a large dynamic range and very fast radiometer designed to precisely observe the meteor light curves. A small series production of a finalized version of the current prototype should accompany some video cameras from the FRIPON network.

## 1 Introduction

The development of a high speed and large dynamic range meteor radiometer was decided for the following main reasons:

- It would be interesting to search for potential correlations between the intriguing oscillations detected by the FRIPON radio network (Rault et al., 2018) appearing sometimes superimposed on the classical smooth meteor head echoes Doppler shifts curves (Figure 1),
- The FRIPON program (Colas et al., 2015) is in need of more accurate and detailed meteor light curves measurements than those obtained with its video cameras network (12 bits & 30 fps devices)

Because the present radiometer prototype is still in an active development phase, this paper is aiming more to give some hints and tricks about the design of the system, and to share the experience gained during the preliminary field tests, rather than to give a detailed and reproducible description of the equipment.

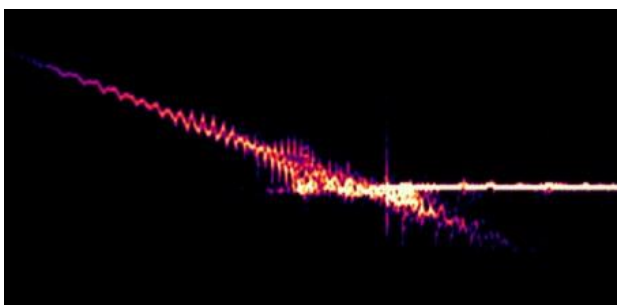

*Figure 1* – Example of some intriguing pulsations observed on a head echo Doppler shift curve

## 2 System design

**State of the art in terms of simple PIN diodes meteor radiometers**

Vida et al. (2015) first proposed a low cost single PIN photodiode meteor radiometer, fitted with a 7.7 mm$^2$ active surface sensor, whose sensitivity was rather low. Then, Segon et al. (2018) obtained encouraging results on bright meteors with a similar system, but equipped with 9 PIN photodiodes, increasing the surface of the light sensor to 67.5 mm$^2$. At last, Buchan et al. (2018) proposed a large dynamic range radiometer using a single large surface light sensor (100 mm$^2$), and a non linear amplification chain allowing a huge dynamic range. Their prototype is still under development, and no meteor detections have yet been reported. The bandwidth offered by these 3 systems is a few hundred Hz.

**Requirements specification for the present radiometer**

The main requirements were as follows:

- High dynamic range (> 12 bits ADC -analog to digital converter-), with magnitudes brighter than -10, which is the FRIPON cameras present limit
- Large bandwidth (several kHz, compared to the 15 Hz limit for the FRIPON cameras running at 30 frames per second )
- Fisheye type sensor FOV (field of view)
- PIN silicium photodiodes sensors technology, to avoid photomultiplier tube systems complexity
- Off-the-shelf recording and data processing softwares
- Accurate data time stamping, to allow fine correlations between video cameras and radiometer data
- Associated PC using Windows 7 or 10 as operating system (but Linux compatible for further development)
- Reasonable cost

**Main technological choices**

A configuration using 16 cheap BPW34 photodiodes was chosen, giving a total 120 mm$^2$ light sensor surface.

Figures 2 and 3 show the relative spectral sensitivity curve and the field of view of such sensors.

At the beginning of the project, it was envisaged to install the 16 sensors on a portion of sphere, in order to create a kind of fly eye allowing a good optical sensitivity at low elevations. This idea was abandoned quickly because in Western Europe, the light pollution radiated by cities during night time is too often annoying. Installing the light sensors on a flat surface as shown on Figure 4 was finally a good choice, as confirmed during the field tests.

The photodiodes are used in the "current mode" instead of "photovoltaic mode" because it allows wider bandwidths Transimpedance amplifiers are used to convert the diodes output currents into voltages (Johnson, 2004). The bandwidth limitation due to the parasitic capacitance of each photodiode is mitigated by the choice of 4 transimpedance amplifiers connected to batches of 4 photodiodes wired in parallel, as shown on Figure 5.

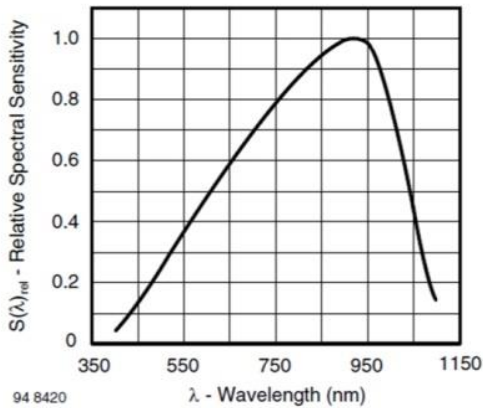

*Figure 2* – Relative spectral sensitivity vs wavelength of the BPW34 photodiode

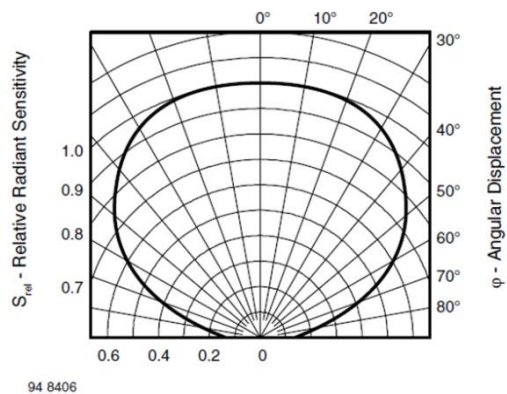

*Figure 3* – Relative radiant sensitivity vs angular displacement of the BPW34 photodiode

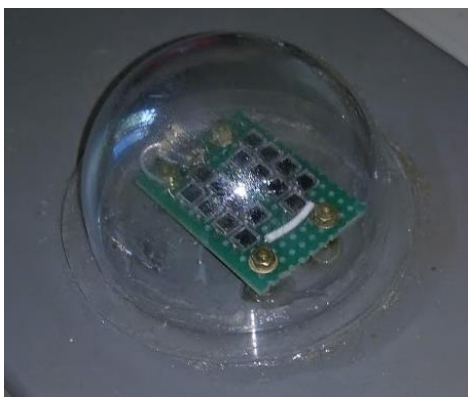

*Figure 4* – Matrix of 4x4 BPW34 photodiodes

The outputs of the transimpedance amplifiers are summed by an operational amplifier.

A TL072 JFET type of operational amplifier was selected because of its large bandwidth, its high linearity, its low noise and its low price.

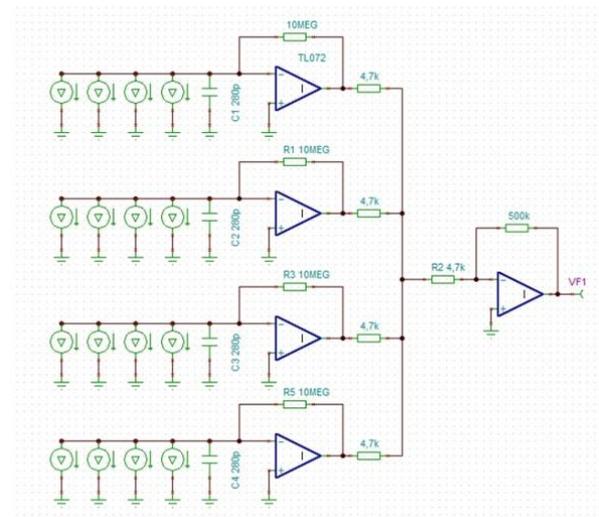

*Figure 5* – Simplified diagram of the analog part of the radiometer

The bandwidth of the entire analog chain (including the photodiodes) was evaluated by means of TINA-Ti v9, a Texas Instruments simulation tool. After carefully adjusting the value of the feedback capacitors installed on each transimpedance amplifier, the bandwidth @ -3 dB goes from 0 to 20 kHz, as shown on Figure 6

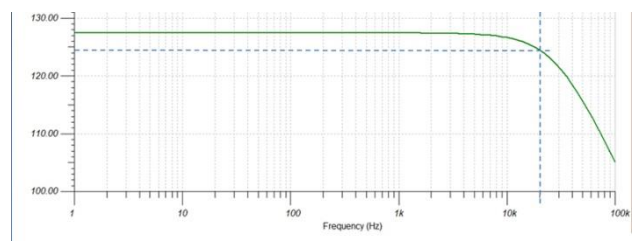

*Figure 6* – Bandwidth of the analog chain computed by the TINA-Ti v9 simulation tool

A DI-1120 (from DATAQ Instruments) 4 channels / 180 kilo samples per second data acquisition system with a resolution of 14 bits was selected for the present prototype. It is connected to the computer via an USB link.

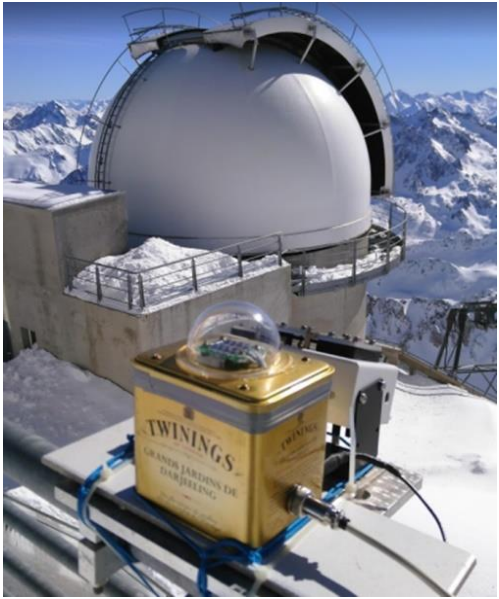

*Figure 7* – "Low noise" configuration

Two main configurations were tested for this prototype of radiometer:

- A very low noise configuration, using the photodiodes and the analog amplification chain only, embedded in a tea box (Figure 7)
- An "all in one" configuration, including the photodiodes, the analog amplification chain, the analog to digital converter, a USB to RJ45 Ethernet converter and a DC/DC 5V / + and – 12V switching power supply embedded in a ruggedized steel box (Figure 8).

The ADC is used in a "two channels" mode, each channel being sampled at 40 kHz. The first channel is connected to a GPS output delivering accurate 1 PPS (pulse per second) for a precise data time stamping purpose. The second channel is used to record the light curves observed by the photometer.

The second version of the device (called "all in one") is easier to use on the field, because there is only one single Ethernet cable connecting the outdoors device to the indoors computer. However, as already mentioned by Segon et al. (2018) and by Buchan et al. (2018), the radiation of electromagnetic interferences by the digital part to the analog chain of the system creates internal noises on the very weak light curves. So the first configuration, although its requires two separate cables (one for the desired light curves signals, another one for the + and -12 V power supply), is preferable to avoid any interferences radiated by the digital parts of the photometer.

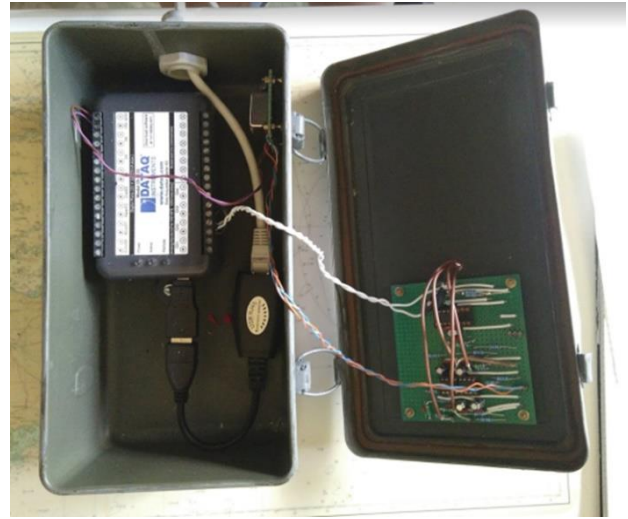

*Figure 8* – "All in one" configuration

The light curves analog data are converted by the DATAQ ADC into digital data that are recorded on a hard disk or an SD memory card in a manufacturer's proprietary WDH format. These WDH format files can be replayed and analyzed thanks to the WINDAQ Browser software suite offered by DATAQ. However, to improve the quality of the observed meteor light curves, the WDH files can be easily translated into classical WAV files for further data processing. It is then possible to perform filtering and signal to noise improvement functions thanks to various well known off-the-shelf audio processing software such as Audacity, Adobe Audition or IzoTope RX7.

## 3 Preliminary observation results

Several observations campaigns during different meteor showers have been performed to date in various locations, such as Observatoire du Pic du Midi, Observatoire de Haute-Provence, but also at home (in a high urban light pollution area) and in a remote alpine chalet located in a remote valley of the Jura Mountains (providing an excellent dark sky location).

**Tentative taxonomy of some night time phenomena**

The most significant initial finding was that observing the night skies in Western Europe with a sensitive and high speed photometer produces a large quantity of anthropic artefacts. Airplanes navigation and green LIDAR flashes (Figure 9), distant wind turbines flashes reflected by high altitude fluctuating clouds (Figure 10), flashes from groomers preparing the ski slopes of a distant ski station (Figure 11), etc. are very common as soon the Sun goes down …

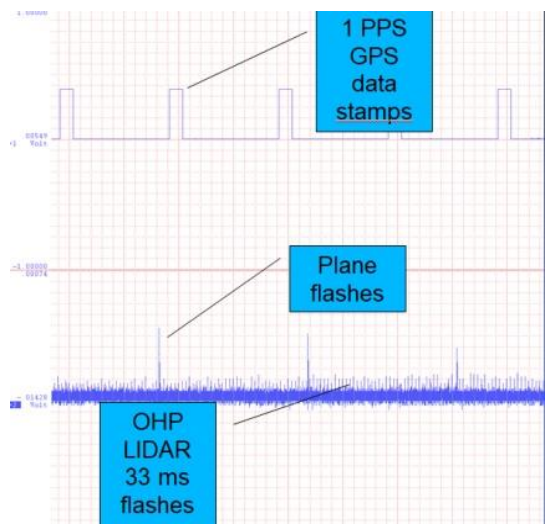

*Figure 9* – Airplanes navigation flashes and green light LIDAR as seen at Observatoire de Haute-Provence

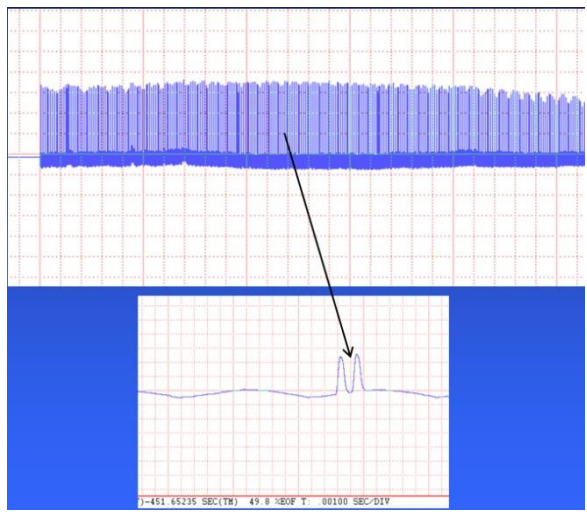

*Figure 10* – Distant wind turbines red flashes (below the horizon) as detected by reflection on some passing clouds

Natural artefacts such as the Moon light (Figure 12) and distant lightning (Figure 13) are also very frequent.

All these false alarms can be easily identified, but the signature of some distant car headlights or of pedestrians walking by night with a headlamp can be confused with a meteor light curve signature.

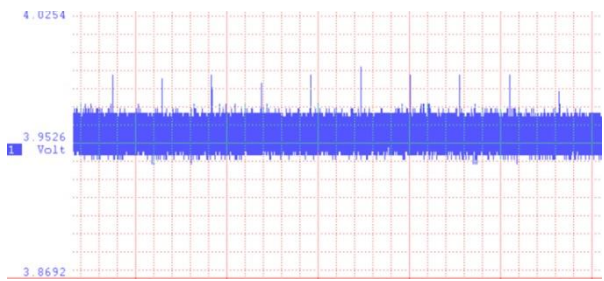

*Figure 11* – Snow groomers at work in the La Mongie ski resort

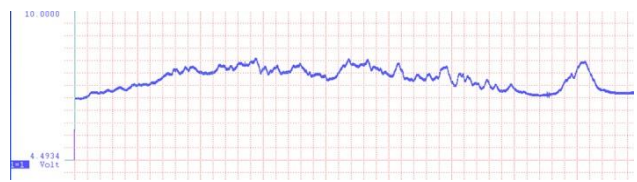

*Figure 12* – Example of Moon light (mag -12.2) modulated by passing clouds

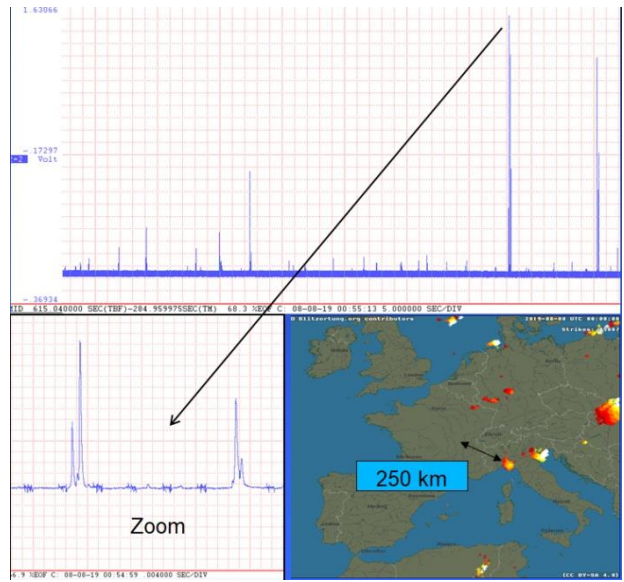

*Figure 13* – Distant lightning over Italy detected in Jura Mountains at a distance of 250 km

The conclusion is that such a photometer must be used in conjunction with the data given by a meteor cameras network, allowing to sort reliably the real meteor light curves and the interferences.

**Improving the signal to noise ratio of the meteor recorded light curves**

Various filters have been successfully tested on the data records to improve the faint meteor light curves signal to noise ratio. Power grid parasitic lines at 50 Hz and their harmonic frequencies at 100 Hz, 150 Hz etc. radiated by the LED and the sodium street lights can be minimized by using a comb filter as shown on Figure 14. Figure 15 shows such a meteor light curve before and after applying the comb filter processing.

The quality of some light curves polluted by complex stationary noises can be enhanced (Figure 16) by using a spectral de-noising function: a copy of the stationary noise (with no meteor light curve) is first analysed by the audio processing software, and then removed from the record containing the meteor light curve.

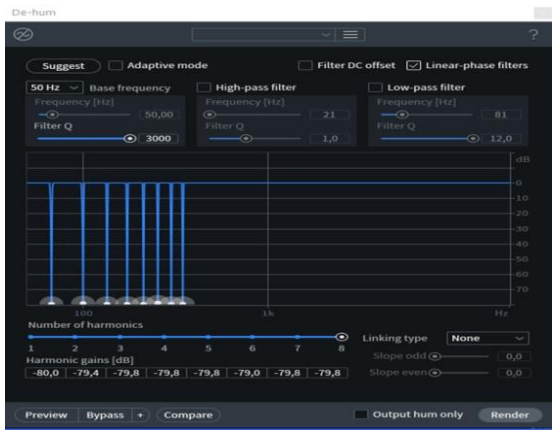

*Figure 14* – Adjustable comb filter offered by the RX7 audio processing software

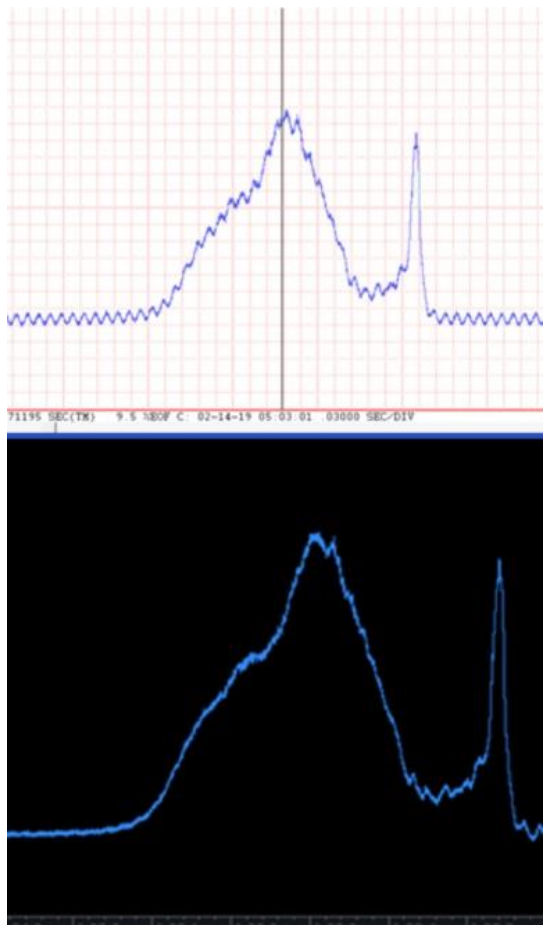

*Figure 15* – Upper curve: raw signal of a meteor light curve. Lower curve: useful signal cleaned by the comb filter

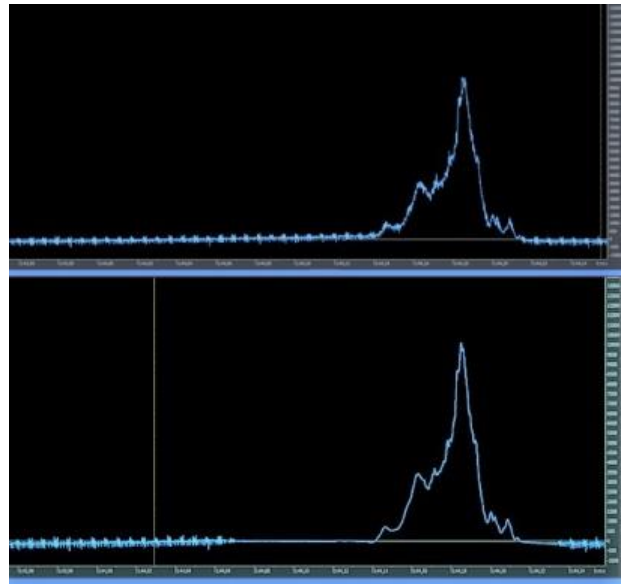

*Figure 16* – Upper curve: raw signal of a meteor light curve polluted by a complex stationary noise. Lower curve: useful signal cleaned by the de-noising function of the RX7 audio processing software

## 4 Discussion

The present radiometer is still in development, mainly to improve as much as possible the signal to noise ratio on the low intensity meteor lights curves. The main improvement track that is still being explored is to find the most effective isolation between the analog and digital parts of the system. In the "low noise" configuration, a transmission of the signals to the indoors ADC via a shielded differential line instead of a coaxial cable is considered.

## 5 Conclusion

The present little tour across the wonderful realm of meteor radiometry clearly shows that the road is long and full of obstacles when aiming at good quality light curves observations. The natural and anthropic false alarms and the internal noises created by the digital components radiating spurious noises into the high gain analog chain of the radiometer forces the user to follow different tactics. In summary:

- During the development phase of a photometer, separate or shield appropriately the analog and the digital parts of the equipment, and use low noise power supplies such as batteries (no switching mode DC converters, etc.)
- During the observation campaigns, choose a location that is protected as well as possible from any light pollution such as cities lights, roads, farms, etc.
- During the phase of data reduction, always use some data obtained by a meteor cameras network, to correlate the presence of a real meteor with the potential light curve detected by the radiometer. Be cautious when using digital processing such as filtering (high pass, low pass, notch filters), or spectral de-noise functions.

These functions have to be carefully adjusted to decrease the interferences without distorting the useful signals, i.e. the desired real meteor light curves.

## Acknowledgement

Many thanks to Denis Vida (Vida et al., 2015), Renato Turcinov (Segon et al., 2018), Hadrien Devillepoix (Buchan et al., 2018) and David Darson (Ecole Normale Supérieure, Paris, France) for our fruitful dialogues about the art of meteor radiometry.